# Imaging 3D Printed Fracture Networks under Stress using X-Ray Computed Tomography

**A. Patsoukis Dimou[1,4], Q. Lei[2], N. Watanabe[3] and A. Suzuki[1]***

[1]Institute of Fluid Science, Tohoku University, Sendai, Japan.

[2]Department of Earth Sciences, Uppsala University, Upppsala, Sweden.

[3]Graduate School of Environmental Studies, Tohoku University, Sendai, Japan.

[4]School of Physics and Astronomy, Edinburgh University, Edinburgh, United Kingdom.

Corresponding author: Anna Suzuki (anna.suzuki@tohoku.ac.jp)

## Key Points:

- We present a novel experimental approach to image 3D fracture network deformation using 3D printing and X-ray CT imaging.
- The experimental results are validated through a comparison with analytical solutions and numerical simulations.
- Our experimental methodology opens a new avenue for aperture characterization from single fractures to complex fracture networks.

## Abstract

We present a novel experimental approach based on 3D printing and X-ray computed tomography to characterize fracture aperture distribution and evolution in 3D fracture networks under varying stress loa–ding conditions. We validate our methodology by comparing experimentally measured stress-dependent fracture apertures with both analytical solutions and numerical simulations, for both single fracture and fracture network scenarios. We show that, for fracture deformation under linear elastic regime, our experimental results agree with numerical simulation. Furthermore, by combining our new experimental methodology with advanced numerical simulations, we illuminate the complex interplay of stress and topology in fracture network deformation. Our approach opens the door for experimentally observing deformable fracture networks under stress in 3D space, which has significant implications for understanding and predicting many geophysical problems involving fractured geological media.

## Plain Language Summary

Fractures are essential pathways for fluid to flow through subsurface geological formations. Understanding how fracture networks behave under in-situ stress conditions is crucial for a variety of underground engineering applications including $CO_2$ sequestration, hydrogen storage, deep mining, geothermal energy, and nuclear waste disposal. Despite extensive studies on numerically modeling stress-dependent deformation in fractured rocks, limited research has been done in the laboratory due to the limitations of conventional experimental approaches such as difficulty in characterizing and visuallizing fractures at sufficient resolution. In this work, we use 3D printing technology to generate customerized fracture networks with full control. Furthermore, we use X-Ray CT imaging to visualize how the fracture network deforms under various stress loading conditions. We finally compare the experimental results with analytical solutions and numerical models which show an excellent agreement. Our experimental methodology can be used with high confidence to investigate fracture network deformation under stress conditions for a range of geophysical applications involving fractured geological media.

## 1 Introduction

Characterizing fractures in geological media is of central importance for many subsurface engineering problems such as carbon sequestration (Bachu, 2008), hydrogen storage (Heinemann et al., 2021), geothermal energy (Barbier, 2002), and nuclear waste disposal (Bear et al., 2012). Here, fractures often serve as the preferential pathways for subsurface fluid flow (Berkowitz, 2002; Tsang & Neretnieks, 1998), which is strongly affected by in-situ stresses due to mechanisms such as compression-induced closure and shear-induced dilation of fracture apertures (Min et al. 2004; Lei et al. 2015a, 2015b, 2020). To characterize and understand the important stress-dependent deformational responses of fracture networks, extensive analytical (Bai et al., 1999; Lei & Gao, 2018; Oda, 1986; Pollard & Segall, 1987; Sneddon & Mott, 1997) and numerical (Banks-Sills, 1991; Davy et al., 2018; Lang et al., 2018) models have been developed. Although these models provide valuable insights, their validation against experimental data remains challenging, due to the inaccessibility of fracture systems in the field and the difficulty of reconstructing fracture networks in the laboratory.

Conventionally, in order to experimentally characterize deformable fracture networks, rock samples are extracted from a target site, with fracture apertures back-calculated from fluid flow experiments (Silliman, 1989). Recent advances in X-ray imaging enable visualizing the internal structure of fractured rocks (Cartwright-Taylor et al., 2022; Montemagno & Pyrak-Nolte, 1995; Pyrak-Nolte et al., 1997), which is however faced with major difficulties. First, it is difficult to precisely fully control fracture network characteristics during sample preparation (Renard et al., 2019). Second, natural rock samples are characterized by multiple scales of heterogeneity that may affect the sample deformation and flow properties (Van Stappen et al., 2022; Liu et al., 2017), complicating the characterization of fracture network behavior.

Additive manufacturing or 3D printing offers a promising solution to overcome this challenge. 3D printing technology enables the cost-effective fabrication of complex 3D geometries with precise control over the fracture network characteristics, allowing detailed investigation of stress impacts on designed fracture systems. 3D printed devices have been validated for their use for flow experiments (Konno et al., 2023; Patoukis Dimou et al., 2021; Patsoukis Dimou et al., 2022; Suzuki et al., 2019) as well as for the investigation of the mechanical tests under static and dynamic loading conditions (Zhou et al., 2019; Zhu et al., 2018). However, the applicability of 3D-printed fracture networks for studying fracture network deformation under varying stress conditions remains largely unexplored. In this work, we aim to, for the first time, experimentally observe at high resolution fracture network deformation under stress loading. We focus on the linear elastic regime to demonstrate the approach. By leveraging the recent development of 3D printing technology in conjunction with X-ray CT imaging, we provide a novel methodological framework for not only experimentally observing fracture network behavior but also examining theoretical and numerical models in predicting fracture network deformation under stress.

## 2 Materials and Methods

### 2.1 Sample Preparation

Fracture geometries were designed using the OpenSCAD software (OpenSCAD, 2021) (Figure 1). The fractures are penny-shaped having an ellipsoidal profile with a radius of $r$ = 10 mm and a maximum aperture of $a$ = 0.5 mm. The first model (denoted as SF1 indicating the first single fracture case) consists of a single penny-shaped fracture horizontally placed. The second model (denoted as SF2) consists of a single penny-shaped fracture with an inclination angle of 30°. The third model (denoted as FN1, indicating the first fracture network case) consists of three fractures, marked as I, II, and III, all with an inclination angle of 30°. The last model (denoted as FN2) superimposes four additional fractures into the system of FN1, forming an intersected fracture network. The samples were prepared as cylinders measuring 3 cm in both height and diameter.

After the digital design models were developed, they were imported to AGILISTA 3100, a Material Jetting 3D printer, for fabrication. The printer uses two types of resins, i.e., AR-M2 and AR-S1. The AR-M2 is deposited in the solid regions, while the AR-S1 is deposited in the flow path regions for structural support. The AR-S1 is water soluble and is dissolved after the printing by immersing the 3D printed product in distilled water. Therefore, a narrow vertical tube with a diameter of 1 mm was preset at the center of each sample, intersecting with fractures and allowing to drain the water-dissolvable support material temporarily deposited within the aperture space during the printing.

The material AR-M2 has a Young's modulus of $E$ = 2.2-2.4 MPa, a Poisson ratio of $\nu$ = 0.3 and a density of $\rho$ = 1,104 kg.m$^{-3}$, provided by KEYENCE Co. The printer's resolution is 15 μm along the X, Y and Z axes.

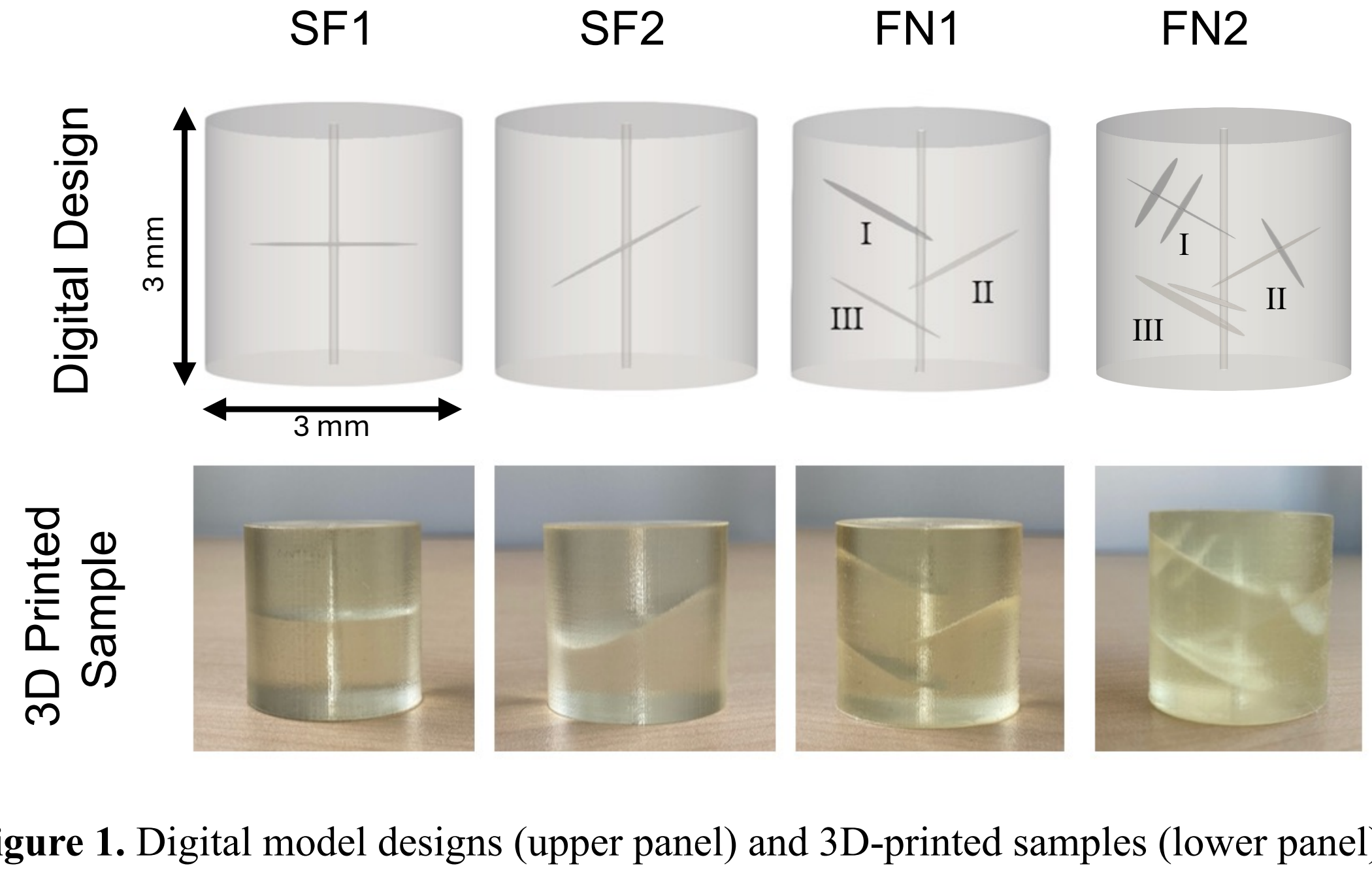

**Figure 1.** Digital model designs (upper panel) and 3D-printed samples (lower panel) for single fracture scenarios (SF1 and SF2) and fracture network scenarios (FN1 and FN2).

## 2.2 Experimental Procedure

After the fracture network geometries were 3D printed, they were placed into a pressure vessel, sealed, and subjected to uniaxial loading. The design of the pressure vessel can be seen in Figure 2. The sample was firstly inserted between two cylinders with a diameter of 3 cm and held in place using a gum sleeve. The cylindrical specimen was then placed in a cell and securely sealed. A hydraulic pump (Intelligent Pump 301) was connected via steel tubing to a pocket located above the top of the cylinder. Furthermore, a pressure gauge was connected to the pocket to measure the pressure directly inside the vessel. The hydraulic pump was then used to inject ethanol to increase the pressure inside the pocket and to subsequently apply load at the top surface of the 3D-printed specimen. After the pressure has reached the required value, the injection stopped with the valve closed, and the experimental cell was disconnected from the hydraulic pump and inserted in an X-ray CT apparatus (ScanXmate-D225RSS27, Gomscan Co., Ltd) to image the internal structure. The samples were scanned using a voltage of 130 kV and a target current of 220 μA, over 3000 projections. The high-resolution of X-Ray CT imaging was conducted at 25 microns per voxel, enabling visualization of changes in the internal structure. The applied stress levels at the specimen were $S$ = 0, 5, and 9.5 MPa. To enable accurate comparison with analytical/numerical solutions (see Section 2.4), a vertical reference plane was defined, passing through the centre of the sample and intersecting the centre of all fractures. The aperture profile was measured along the central plane of the specimen, as shown in Figure 2b and 2c.

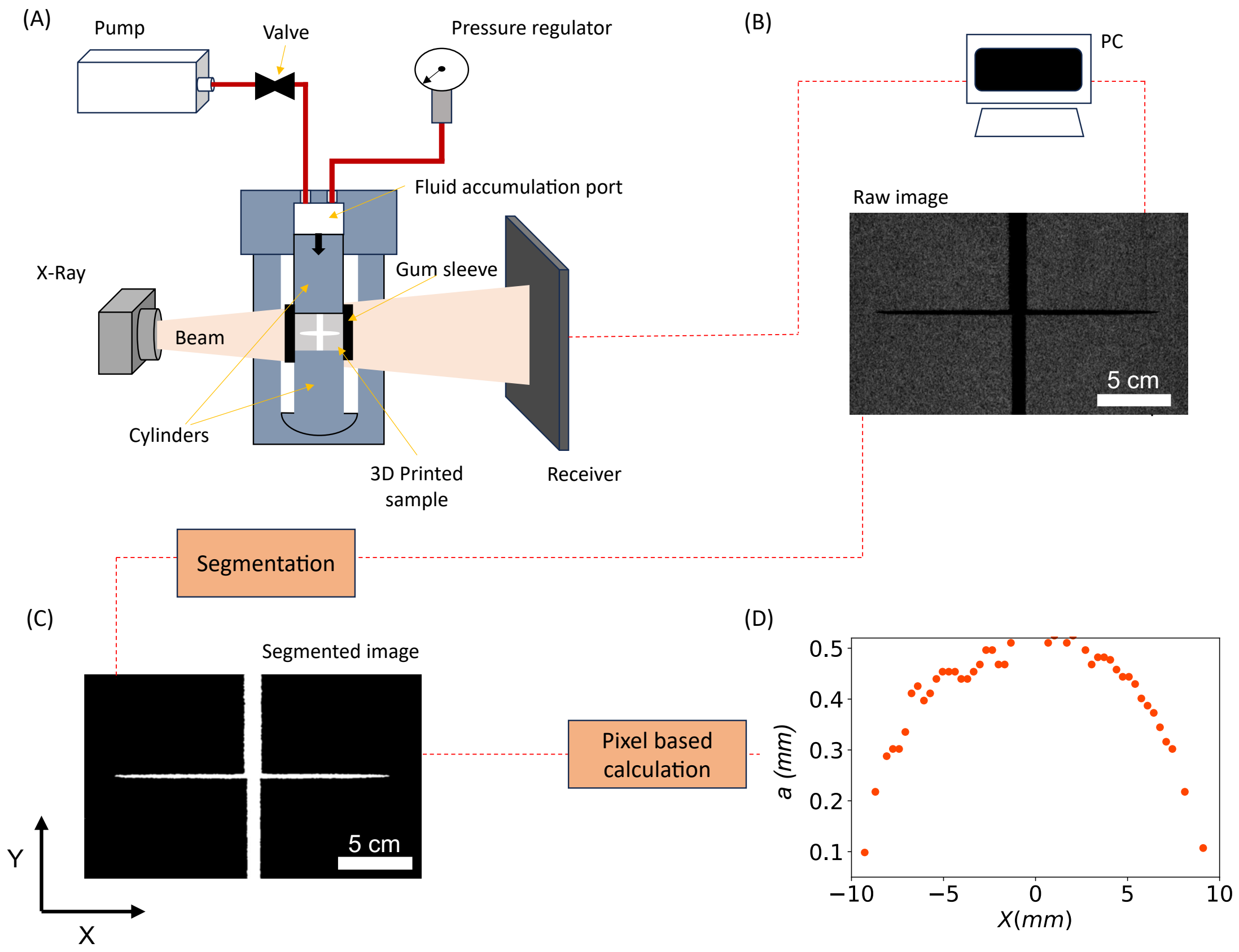

**Figure 2.** (a) Experimental apparatus with pressure vessel, hydraulic pump, pressure regulator, and X-Ray unit. (b) Raw image recorded by the X-Ray imaging system. (c) Segmented image of solid (black) and fracture (white). (d) Derived fracture aperture profile along the X axis.

## 2.3 Image Processing

After completing the X-ray CT experiments, the images were analyzed using Python and Fiji software (Schindelin et al., 2012). First, the images were denoised using the contrast-limited adaptive histogram equalization (Zuiderveld, 1994), followed by an adaptive Wiener filter. The image was then segmented into solid and fracture space using the Trainable Weka Segmentation algorithm (Arganda-Carreras et al., 2017). Finally, a Python script was employed to analyze the fracture aperture using a pixel-based calculation method. The analysis is performed within the reference plane as described in Section 2.1. The raw image and the segmented image for the single fracture model SF1 are shown in Figure 2B and 2C for illustration.

## 2.4 Analytical and Numerical Models

The aperture $a$ (m) of a penny-shaped fracture experiencing linear elastic deformation under uniaxial stress loading can be calculated analytically (Sneddon, 1946). The analytical model assumes a penny-shaped fracture in an infinite, homogeneous domain, with the solution given in a dimensionless form as follow (see Text S1 in the Supporing Information):

$$\tilde{a}\,(\tilde{r}) = 1 - \frac{16(1-\nu^2)}{\pi^2}\tilde{\sigma}_v\sqrt{1-\tilde{r}^2}, \tag{1}$$

where $\tilde{a}$ (-) is the normalised aperture defined as $\tilde{a} = a/a_0$, $\tilde{r}$ is the dimensionless radial coordinate defined as $\tilde{r} = r/R$, $\nu$ is the Poisson ratio, and $\tilde{\sigma}_v$ is the dimensionless stress defined as $\tilde{\sigma}_v = \sigma_v/K a_0$.

In addition, we perform finite element numerical simulation based on COMSOL Multiphysics software (COMSOL, 2022) to compute fracture network deformation under stress. More details about the numerical model setup can also be found in the Supporting Information. The values of axial stresses used in the numerical simulation were the same as those applied in the experiment, i.e., $S$ = 0, 5 and 9.5 MPa. The bottom boundary was set as a roller condition , while the lateral boundaries have a free-surface condition. These settings reflect the uniaxial compression test condition explored in the laboratory experiment (see Section 2.2). Finally, we compute the difference between the experimental results and the analytical/numerical solutions using:

$$\varepsilon = \frac{1}{N}\sum_{i=1}^{N}\left|a_i^{\text{model}} - a_i^{\text{exp}}\right| \tag{2}$$

where $a_i^{\text{exp}}$ is the aperture of the fracture measured experimentally at position $i$, $a_i^{\text{model}}$ is the aperture of the fracture calculated analytical or numerically at the same position and $N$ is the number of points.

## 3 Results

In this section, the experimental results obtained are compared with the analytical solution and numerical simulation.

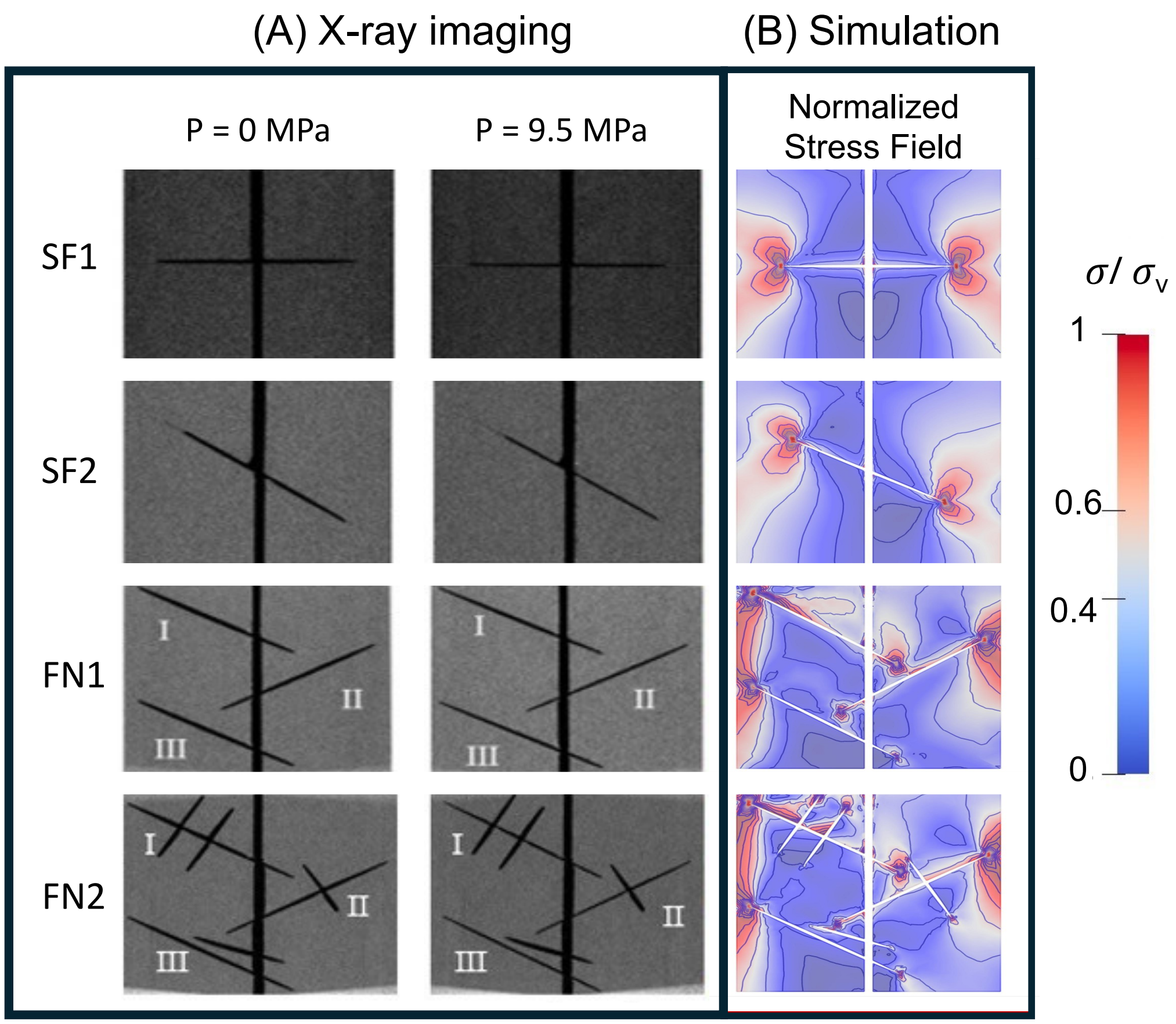

**Figure 3.** (A) X-ray experimental observation of fracture aperture profiles in the 3D printed samples under uniaxial stress loading of $S$ = 0 MPa and 9.5 MPa. Note that the vertical line in the center of the sample is the water injection pipe needed for emptying the fracture from the water dissolvable support. (B) Numerical simulation of stress distribution in the fractured samples.

For clarity, in the Main Text, the experimental results for the stress levels $S$ = 0 and 9.5 MPa are shown only, while the experimental results for the stress $S$ = 5 MPa are provided in the Supporting Information.

Figure 3A presents the X-Ray CT images for $S$ = 0 and 9.5 MPa, where the internal structure of fractures and fracture networks are visualized at a high resolution. By examining the X-ray images at 0 MPa, we observe that the fracture has its maximum aperture at the center, gradually narrowing toward the tips, consistent with the ellipsoidal profile of the original design. Some artificial imperfections are visible in the sample, likely resulting from 3D printing errors. Furthermore, it can be seen that with an increased stress loading, the fractures apertures decrease in all cases.

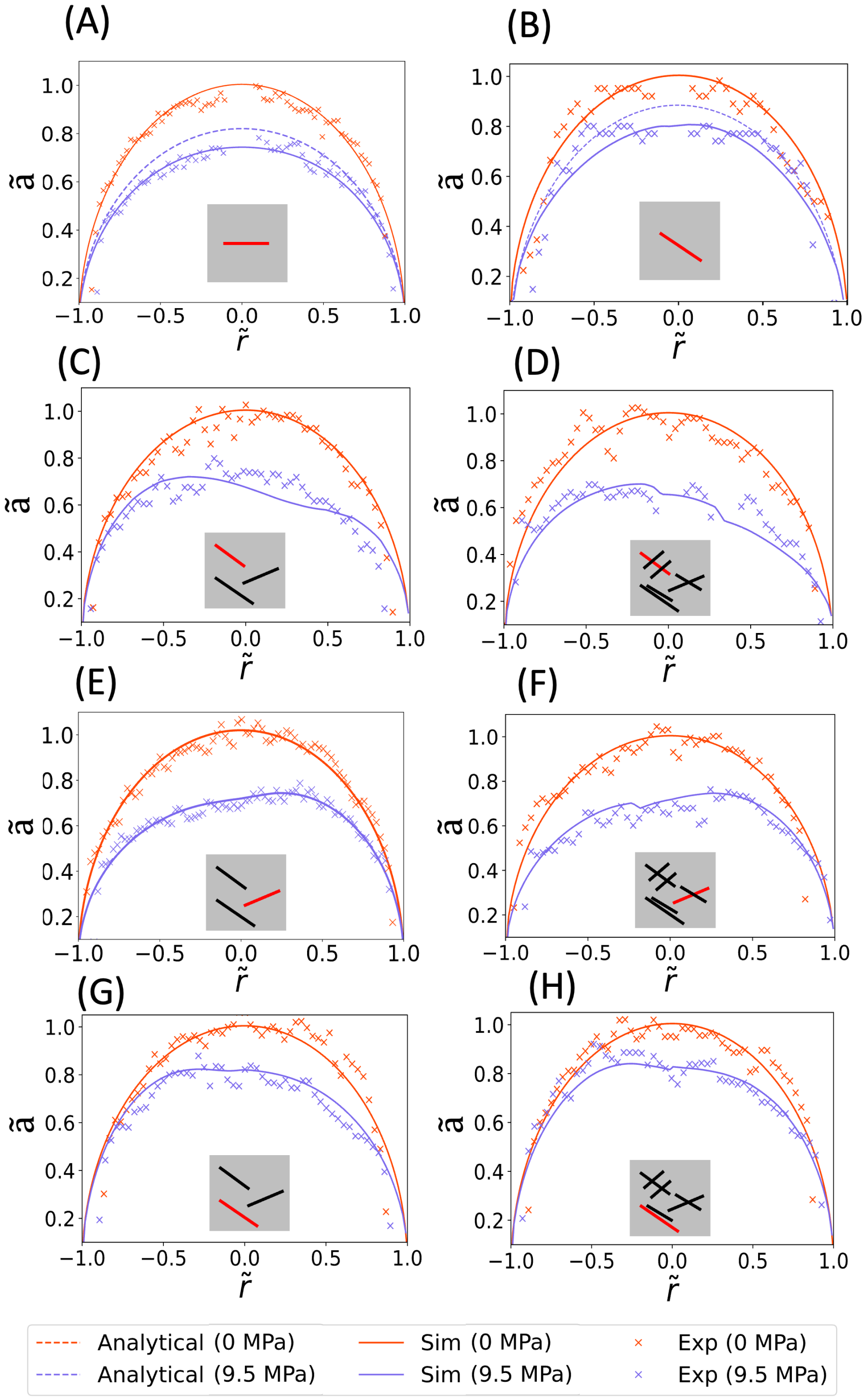

**Figure 4.** Comparison of the fracture appertures predicted between experimental measurements, numerical simulations, and analytical solutions for the cases of SF1, SF2, FN1 and FN2. The analytical solutions are only available for the single fracture scenarios

SF1 and SF2. (A) SF1, (B) SF2, (C) $FN1_I$, (D) $FN2_I$, (E) $FN1_{II}$ (F) $FN2_{II}$, (G) $FN1_{III}$, (H) $FN2_{III}$. Subscripts I, II, and III denote the fractures illustrated in Figure 1.

Figure 3B shows the stress field captured in the numerical simulation for stress loading $S$ = 9.5 MPa. Figure 4 displays the aperture profiles obtained from the analytical, numerical, and experimental models. For the SF1 model, as can be seen in Figure 3B, the stress field is symmetrically distributed with respect to the fracture center, where enhanced stress concentrations locally occur at the fracture tips. Furthermore, from Figure 4A, one can see that for $S$ = 9.5 MPa, the numerical simulation, analytical solution, and experiment data show good agreement. We repeated the same experiment two times and the results were consistent. More specifically, comparing the numerical simulation with the experimental results, we find a good agreement with an overall error of $\varepsilon_{SF1_1}^{0MPa} = 3.93\%$ and $\varepsilon_{SF1_1}^{9.5MPa} = 3.78\%$ for the first experiment, $\varepsilon_{SF1_2}^{0MPa} = 3.68\%$ and $\varepsilon_{SF1_2}^{9.5MPa} = 3.85\%$ for the second experiment, suggesting our experimental method and the results are robust. Furthermore, this agreement confirms our assumption that the 3D printed specimens follow linear elastic deformation under the given stress loading. The experimental error remains at the same values $\varepsilon < 5\%$ and does not change under different stress conditions, which can be attributed to systemic errors related to 3D printing or image analysis/segmentation. Furthermore, Figure 4 shows that the experimental results neither overestimate nor underestimate the fracture aperture. This further indicates that the error could be due to slightly irregular surfaces or segmentation errors. When comparing the results obtained from the analytical equation with the numerical simulation, it is observed that there is a difference in the fracture aperture calculated by the two methods. The cause of this error is considered to be due to boundary effects. Since the analytical solution assumes an infinite medium, it does not take account of the impact of boundaries on the stress distribution in the laboratory specimen. This is further investigated and confirmed in the Supporting Information, which shows that when the radius of the outer boundary of a specimen is larger than 30 cm (i.e., ten times the actual specimen used in the experiment), the analytical solution becomes to agree well with the numerical simulation.

For the model SF2, in Figure 3B we can see that the stress distribution within the specimen, where the fracture has a 30-degree inclination and stress concentrations mainly occur at the fracture tips. It seems that the change in the stress distribution (compared to SF1) within the specimen has a minor effect on the stress distribution at the fracture surface and therefore small influence on the aperture profile. When comparing the experimental and numerical results, seen in Figure 4B, there is a good agreement with an overall error of $\varepsilon_{SF2}^{0MPa} = 4.76\%$ and $\varepsilon_{SF2}^{9.5MPa} = 4.88\%$, which further demonstrates the accuracy and robustness of our experimental methodology.

For the model FN1, the co-existence of three fractures in the sample leads to a more complex stress distribution within the specimen. Figure 3 shows that the repositioning of the fractures inside the specimen and the existence of neighboring fractures both contribute to the emergence of an asymmetrical stress distribution along the fractures. In Figure 4C 4D, and 4E, it can be seen that this irregular distribution of stress leads to an asymmetrical fracture closure behavior. Comparing the experimental data with the numerical simulation, we find that the fracture aperture profiles derived by the two methods are in a good agreement with an overall error $\varepsilon < 5\%$, similar to the single fracture cases.

For the model FN2, as can be seen in Figure 3B, adding more fractures to the system (FN2) leads to further redistribution of the stress field. Moreover, we can see the strong stress perturbation due to the introduction of new fractures that intersect with those fractures existing in FN1. This difference in the redistribution of stresses locally at the fracture surfaces is larger when comparing fracture I (shared by FN1 and FN2) that is crossed by two new fractures in FN2 (see Figure 4C and 4D). The difference is smaller for fracture II that is crossed by one additional fracture in FN2 (see Figure 4E and 4F). Even smaller difference is observed for fracture III where a new fracture is placed close to but not crossing it in FN2 (see Figure 4G and 4H). This reflects the important role of fracture interaction on the stress distribution in fracture networks, which has been well captured in both our experimental observations and numerical models.

## 4 Discussion

In this study, we present an experimental methodology that enables the visualization of complex fracture network deformation in 3D by combining 3D printing technology with X-ray CT imaging. This approach benefits from the ability to precisely control fracture geometry using 3D printing and to visualize internal structures in detail using X-ray imaging. Together, these techniques allow for the generation of fracture networks with defined geometrical properties, such as inclination and connectivity, and for the direct observation and quantification of fracture apertures under various stress conditions. While it is often difficult to precisely characterize fractures in natural rocks (F. Van Stappen et al., 2022), limiting systematic investigation of fracture network behavior, our approach offers a reproducible and flexible framework to study this problem under controlled conditions. Laboratory experiments inherently suffer from finite size effects due to constraints imposed by the testing equipment, logistics, or the limited dimensions of samples, which are often extracted from boreholes. By integrating laboratory experiments, numerical simulations, and analytical solutions, our approach provides a unified workflow for investigating fracture network behavior, where numerical models can be validated against experimental data with the boundary effects considered. Simultaneously, these numerical models with an enlarged domain setting can be verified against analytical solutions that assume infinite domains. This dual process of validation and verification ensures that the numerical models are robust across both idealized and practical conditions. Consequently, they can be reliably applied to explore real-world geological systems, which typically lie between the two end-member cases of infinite domains and finite laboratory samples. By performing experiments using identical geometries, we demonstrate the consistency and repeatability of our experimental methodology. Furthermore, we show that comparing the experimental results obtained with our methodology to the results from the numerical simulations reveal good agreement, validating our experimental approach.

Some limitations remain and warrant further investigation. The experiments presented in this study were conducted within the linear elastic deformation regime, whereas some geological processes may involve inelastic deformation or even fracture propagation. While our 3D printed models reproduce elastic responses with sufficient fidelity for the purpose of validating numerical models in the linear elastic regime, they do not fully represent the mechanical complexity of natural rocks. Additionally, there are potential sources of systematic error, including segmentation uncertainty during image processing and inaccuracies during the 3D printing process. These errors while minor (less than 5%) can be further reduced with advances in 3D printing resolution and image analysis. Additionally, the irregular surfaces produced by 3D printing can be used as inputs for the numerical simulations to account for the natural surface roughness, providing a more accurate comparison between physical and digital geometries. Our experiments have mainly focused on single fracture and relatively simple fracture network geometries, whereas much more complicated fracture network geometries can also be generated using 3D printing (Konno et al., 2023; Suzuki et al., 2017, 2019). Finally, our current experiments were performed under uniaxial loading for the demosntration purpose, yet in natural, systems, triaxial stress conditions may prevail especially for deep subsurface rocks. While this study aimed to validate the methodology under the simple uniaxial loading conditions, the setup can be adapted to combine with triaxial loading facilities in the future work.

Our experimental technique for generating and visualizing fracture deformation under varying stress conditions can also be combined with fluid flow experiments to investigate the hydraulic behavior of fractured media. Such efforts can provide significant insights into the intricate interplay between fracture pattern, stress loading, and fluid flow. Additionally, it can serve as an experimental methodology to validate numerical models that simulate stress effects on fluid flow through fracture networks. This will contribute to the development of more reliable tools for understanding and predicting relevant geophysical processes during industrial activities such asgeothermal energy exploitation, $CO_2$ storage, and hydrogen storage.

# Acknowledgments

This work was supported by the JSPS KAKENHI, Japan [grant numbers JP20H02676, JP22H05108, JP24K01413].

# Conflict of Interest

The authors declare that they have no conflict of interest.

**References**

Arganda-Carreras, I., Kaynig, V., Rueden, C., Eliceiri, K. W., Schindelin, J., Cardona, A., & Sebastian Seung, H. (2017). Trainable Weka Segmentation: a machine learning tool for microscopy pixel classification. *Bioinformatics (Oxford, England)*, *33*(15), 2424—2426. https://doi.org/10.1093/bioinformatics/btx180

Bachu, S. (2008). CO2 storage in geological media: Role, means, status and barriers to deployment. *Progress in Energy and Combustion Science*, *34*(2), 254–273. https://doi.org/https://doi.org/10.1016/j.pecs.2007.10.001

Bai, M., Meng, F., Elsworth, D., & Roegiers, J.-C. (1999). Analysis of Stress-dependent Permeability in Nonorthogonal Flow and Deformation Fields. *Rock Mechanics and Rock Engineering*, *32*(3), 195–219. https://doi.org/10.1007/s006030050032

Banks-Sills, L. (1991). Application of the Finite Element Method to Linear Elastic Fracture Mechanics. *Applied Mechanics Reviews*, *44*(10), 447–461. https://doi.org/10.1115/1.3119488

Barbier, E. (2002). Geothermal energy technology and current status: an overview. *Renewable and Sustainable Energy Reviews*, *6*(1), 3–65. https://doi.org/https://doi.org/10.1016/S1364-0321(02)00002-3

Bear, J., Tsang, C.-F., & De Marsily, G. (2012). *Flow and contaminant transport in fractured rock*. Academic Press.

Berkowitz, B. (2002). Characterizing flow and transport in fractured geological media: A review. *Advances in Water Resources*, *25*(8), 861–884. https://doi.org/https://doi.org/10.1016/S0309-1708(02)00042-8

Cartwright-Taylor, A., Mangriotis, M.-D., Main, I. G., Butler, I. B., Fusseis, F., Ling, M., Andò, E., Curtis, A., Bell, A. F., Crippen, A., Rizzo, R. E., Marti, S., Leung, Derek. D. V, & Magdysyuk, O. V. (2022). Seismic events miss important kinematically governed grain scale mechanisms during shear failure of porous rock. *Nature Communications*, *13*(1), 6169. https://doi.org/10.1038/s41467-022-33855-z

COMSOL. (n.d.). *COMSOL Multiphysics®* (v. 6.1.).

Davy, P., Darcel, C., Le Goc, R., & Mas Ivars, D. (2018). Elastic Properties of Fractured Rock Masses With Frictional Properties and Power Law Fracture Size Distributions. *Journal of Geophysical Research: Solid Earth*, *123*(8), 6521–6539. https://doi.org/https://doi.org/10.1029/2017JB015329

F. Van Stappen, J., McBeck, J. A., Cordonnier, B., Pijnenburg, R. P. J., Renard, F., Spiers, C. J., & Hangx, S. J. T. (2022). 4D Synchrotron X-ray Imaging of Grain Scale Deformation Mechanisms in a Seismogenic Gas Reservoir Sandstone During Axial Compaction. *Rock Mechanics and Rock Engineering*, *55*(8), 4697–4715. https://doi.org/10.1007/s00603-022-02842-7

Heinemann, N., Alcalde, J., Miocic, J. M., Hangx, S. J. T., Kallmeyer, J., Ostertag-Henning, C., Hassanpouryouzband, A., Thaysen, E. M., Strobel, G. J., & Schmidt-Hattenberger, C. (2021). Enabling large-scale hydrogen storage in porous media–the scientific challenges. *Energy & Environmental Science*, *14*(2), 853–864.

Konno, M., Patsoukis Dimou, A., & Suzuki, A. (2023). Using 3D-printed fracture networks to obtain porosity, permeability, and tracer response datasets. *Data in Brief*, *47*, 109010. https://doi.org/https://doi.org/10.1016/j.dib.2023.109010

Lang, P. S., Paluszny, A., Nejati, M., & Zimmerman, R. W. (2018). Relationship Between the Orientation of Maximum Permeability and Intermediate Principal Stress in Fractured Rocks. *Water Resources Research*, *54*(11), 8734–8755. https://doi.org/https://doi.org/10.1029/2018WR023189

Lei, Q., & Gao, K. (2018). Correlation Between Fracture Network Properties and Stress Variability in Geological Media. *Geophysical Research Letters*, *45*(9), 3994–4006. https://doi.org/https://doi.org/10.1002/2018GL077548

Liu, T., Lin, B., & Yang, W. (2017). Impact of matrix–fracture interactions on coal permeability: Model development and analysis. *Fuel*, *207*, 522–532. https://doi.org/https://doi.org/10.1016/j.fuel.2017.06.125

Montemagno, C. D., & Pyrak-Nolte, L. J. (1995). Porosity of natural fracture networks. *Geophysical Research Letters*, *22*(11), 1397–1400. https://doi.org/https://doi.org/10.1029/95GL01098

Oda, M. (1986). An equivalent continuum model for coupled stress and fluid flow analysis in jointed rock masses. *Water Resources Research*, *22*(13), 1845–1856. https://doi.org/https://doi.org/10.1029/WR022i013p01845

OpenSCAD. (2021). *The programmers solid 3D CAD modeller* .

Patoukis Dimou, A., Suzuki, A., Menke, H., Maes, J., & Geiger, S. (2021). 3D printing-based microfluidics for Geosciences . *ICFD* .

Patsoukis Dimou, A., Menke, H. P., & Maes, J. (2022). Benchmarking the Viability of 3D Printed Micromodels for Single Phase Flow Using Particle Image Velocimetry and Direct Numerical Simulations. *Transport in Porous Media*, *141*(2), 279–294. https://doi.org/10.1007/s11242-021-01718-8

Pollard, D. D., & Segall, P. (1987). *8 – THEORETICAL DISPLACEMENTS AND STRESSES NEAR FRACTURES IN ROCK: WITH APPLICATIONS TO FAULTS, JOINTS, VEINS, DIKES, AND SOLUTION SURFACES*. https://api.semanticscholar.org/CorpusID:132641475

Pyrak-Nolte, L. J., Montemagno, C. D., & Nolte, D. D. (1997). Volumetric imaging of aperture distributions in connected fracture networks. *Geophysical Research Letters*, *24*(18), 2343–2346. https://doi.org/https://doi.org/10.1029/97GL02057

Renard, F., McBeck, J., Kandula, N., Cordonnier, B., Meakin, P., & Ben-Zion, Y. (2019). Volumetric and shear processes in crystalline rock approaching faulting. *Proceedings of the National Academy of Sciences*, *116*(33), 16234–16239. https://doi.org/10.1073/pnas.1902994116

Schindelin, J., Arganda-Carreras, I., Frise, E., Kaynig, V., Longair, M., Pietzsch, T., Preibisch, S., Rueden, C., Saalfeld, S., Schmid, B., Tinevez, J.-Y., White, D. J., Hartenstein, V., Eliceiri, K., Tomancak, P., & Cardona, A. (2012). Fiji: an open-source platform for biological-image analysis. *Nature Methods*, *9*(7), 676–682. https://doi.org/10.1038/nmeth.2019

Silliman, S. E. (1989). An interpretation of the difference between aperture estimates derived from hydraulic and tracer tests in a single fracture. *Water Resources Research*, *25*(10), 2275–2283. https://doi.org/https://doi.org/10.1029/WR025i010p02275

Sneddon, I. N. (1946). The distribution of stress in the neighbourhood of a crack in an elastic solid. *Proceedings of the Royal Society of London. Series A. Mathematical and Physical Sciences*, *187*(7), 229–260.

Sneddon, I. N., & Mott, N. F. (1997). The distribution of stress in the neighbourhood of a crack in an elastic solid. *Proceedings of the Royal Society of London. Series A. Mathematical and Physical Sciences*, *187*(1009), 229–260. https://doi.org/10.1098/rspa.1946.0077

Suzuki, A., Minto, J. M., Watanabe, N., Li, K., & Horne, R. N. (2019). Contributions of 3D Printed Fracture Networks to Development of Flow and Transport Models. *Transport in Porous Media*, *129*(2), 485–500. https://doi.org/10.1007/s11242-018-1154-7

Suzuki, A., Watanabe, N., Li, K., & Horne, R. N. (2017). Fracture network created by 3-D printer and its validation using CT images. *Water Resources Research*, *53*(7), 6330–6339. https://doi.org/https://doi.org/10.1002/2017WR021032

Tsang, C.-F., & Neretnieks, I. (1998). Flow channeling in heterogeneous fractured rocks. *Reviews of Geophysics*, *36*(2), 275–298. https://doi.org/https://doi.org/10.1029/97RG03319

Zhao, Z., Jing, L., & Neretnieks, I. (2010). Evaluation of hydrodynamic dispersion parameters in fractured rocks. *Journal of Rock Mechanics and Geotechnical Engineering*, *2*(3), 243–254.

Zhao, Z., Jing, L., Neretnieks, I., & Moreno, L. (2011). Numerical modeling of stress effects on solute transport in fractured rocks. *Computers and Geotechnics*, *38*(2), 113–126.

Zhao, Z., Rutqvist, J., Leung, C., Hokr, M., Liu, Q., Neretnieks, I., Hoch, A., Havlíček, J., Wang, Y., Wang, Z., Wu, Y., & Zimmerman, R. (2013). Impact of stress on solute transport in a fracture network: A comparison study. *Journal of Rock Mechanics and Geotechnical Engineering*, *5*(2), 110–123. https://doi.org/https://doi.org/10.1016/j.jrmge.2013.01.002

Zhou, T., Zhu, J. B., Ju, Y., & Xie, H. P. (2019). Volumetric fracturing behavior of 3D printed artificial rocks containing single and double 3D internal flaws under static uniaxial compression. *Engineering Fracture Mechanics*, *205*, 190–204. https://doi.org/https://doi.org/10.1016/j.engfracmech.2018.11.030

Zhu, J. B., Zhou, T., Liao, Z. Y., Sun, L., Li, X. B., & Chen, R. (2018). Replication of internal defects and investigation of mechanical and fracture behaviour of rock using 3D printing and 3D numerical methods in combination with X-ray computerized tomography. *International Journal of Rock Mechanics and Mining Sciences*, *106*, 198–212. https://doi.org/https://doi.org/10.1016/j.ijrmms.2018.04.022

Zuiderveld, K. J. (1994). Contrast Limited Adaptive Histogram Equalization. *Graphics Gems*. https://api.semanticscholar.org/CorpusID:62707267